\documentclass[prd,preprint,superscriptaddress]{revtex4}
\usepackage{revsymb}
\newcommand{\be}{\begin{equation}}
\newcommand{\ee}{\end{equation}}
\newcommand{\ben}{\begin{eqnarray}}
\newcommand{\een}{\end{eqnarray}}
\newcommand{\n}{\label}
\newcommand{\no}{\noindent}

\begin{document}

\title{Dark energy, dissipation and the coincidence problem}

\author{Luis P. Chimento\footnote{Electronic address: chimento@df.uba.ar}}
\affiliation{Departamento de F\'{\i}sica,
Universidad de Buenos Aires,
1428 Buenos Aires, Argentina}
\author{Alejandro S. Jakubi\footnote{
Electronic address: jakubi@df.uba.ar}}
\affiliation{Departamento de F\'{\i}sica,
Universidad de Buenos Aires,
1428 Buenos Aires, Argentina}
\author{Diego Pav\'on\footnote{Electronic address: diego.pavon@uab.es}}
\affiliation{Departamento de F\'{\i}sica,
Universidad Aut\'onoma de Barcelona,
08193 Bellaterra, Spain}

\begin{abstract}
In a recent paper we showed that a quintessence scalar field plus a
dissipative matter fluid can drive late cosmic accelerated expansion
and simultaneously solve  the coincidence problem \cite{enlarged}.
In this brief report we extend this result to the cases when the
scalar field is replaced either by a Chaplygin gas or a
tachyonic fluid.
\end{abstract}

%\noindent
%PACS numbers: 98.80.Jk

\pacs{98.80.Jk}

\maketitle

The low luminosity  of supernovae type Ia at high redshifts strongly
suggests that our present Universe experiences a period of
accelerated expansion (see e.g., Ref. \cite{saul} and references therein),
something at variance with the long--lived Einstein--de--Sitter
cosmological model \cite{peacock}. This combined with that the
position of the first acoustic peak of the CMB is compatible with a
critical density Universe (i.e., $\Omega = 1$) and that estimations
of the mass density ($\Omega_{m}\sim 0.3$) indicates that in addition
to luminous and dark matter some other component (usually referred to
as ``dark energy") must contribute to the critical density value.
Moreover, the latter component must, on the one hand, entail a
negative pressure to drive the accelerated expansion and, on the
other hand, cluster only weakly so that the structure formation
scenario does not get spoiled \cite{iap}.

In principle, the obvious dark energy candidate should be a small cosmological
constant. However, on the one hand, there are serious theoretical problems
regarding its small value (many orders of magnitude below the one predicted
by any straightforward quantum field theory) and, on the other hand, it
is unable to give a satisfactory answer to the embarrassing question:
``Why  the vacuum and matter energy densities are of the same order
precisely today?" (One should bear in mind that the former remains
constant with expansion while the latter redshifts approximately as
$a^{-3}$). This is the coincidence problem.

To overcome this problem recourse was repeatedly made to a
self--interacting scalar field $\phi$ with equation of state
$p_{\phi} = (\gamma_{\phi} - 1) \rho_{\phi}$, where $\gamma_{\phi}$
is a time-varying quantity restricted to the range $[0,1]$ so that:
(i) $p_{\phi}$ is always  negative, and (ii) its energy density
is much lower than that of matter (and radiation) at early times
but comparable to the latter at recent times \cite{quintessence}.
Thus, the usual strategy was to assume some potential $V(\phi)$
leading to the desired behavior. As shown by Padmanabhan, it is a
straightforward matter to design a suitable potential
\cite{padmanabhan}.

In a recent paper we demonstrated that a mixture of a perfect matter
fluid and quintessence field, interacting with each other just
gravitationally, cannot drive acceleration and simultaneously solve
the coincidence problem. However when the matter fluid is dissipative
enough (i.e., it possesses a sufficiently large bulk viscous pressure
$\pi$), the coincidence problem can be solved (i.e.,
$\rho_{m}/\rho_{\phi}$ tends to some constant of order unity) and the
Universe has a late accelerated expansion irrespective of the assumed
potential $V(\phi)$ (cf. Ref. \cite{enlarged}).  The proof can be
sketched as follows:

The Friedmann equation plus the conservation equations for matter
and  quintessence  in a Friedmann--Robertson--Walker universe
dominated by these two components (non--interacting with one
another), in terms of the density parameters, are
\\
\begin{eqnarray}
1&=&\Omega_{m}  +  \Omega_{\phi} + \Omega_{k} \, , \\
\dot{\Omega} & = & (3\gamma -2) H (\Omega -1) \Omega \, ,\\
\dot{\Omega}_{\phi} & = & [2+(3\gamma -2) \Omega -
3 \gamma_{\phi}]\Omega_{\phi} H \, ,
\label{eq:ekg}
\end{eqnarray}
\\
where $\Omega \equiv \Omega_{m} + \Omega_{\phi}$, and $\gamma$ stands for
the overall baryotropic index
$\gamma = (\gamma_{m} \Omega_{m} + \gamma_{\phi} \Omega_{\phi})/\Omega$,
with $\gamma_{m,\, \phi} \equiv 1 + (p_{m,\, \phi}/\rho_{m,\, \phi})$,
and such that    $1 \leq \gamma_{m} \leq 2$ and $0 \leq \gamma_{\phi} < 1 $
(it should be noted that in general $\gamma_{m}$ and $\gamma_{\phi}$ may
vary with time).

{}From the above equations it is immediately seen that for $\Omega = 1$
Eq. (3) implies that $\dot{\Omega}_{\phi} > 0$. Consequently, at
large times $\Omega_{\phi} \rightarrow 1$ and $\Omega_{m} \rightarrow
0$, i.e., the accelerated expansion ($q \equiv - \ddot{a}/(a H^{2}) < 0
\Longrightarrow \gamma_{\phi} < 2/3$) and the coincidence problem cannot
be solved simultaneously within this approach. Moreover, for the solution
$\Omega = 1$ to be stable the overall baryotropic index must comply
with the upper bound $\gamma < 2/3$ which is uncomfortably low.

However, things fare differently when one assumes the matter fluid
dissipative. Indeed, Eqs. (2) and (3) generalize to
\\
\begin{eqnarray}
\dot{\Omega} & = &  \left[3 \left(\gamma+
\frac{\pi}{\rho}\right) -2\right] H (\Omega -1) \Omega\, ,  \\
\dot{\Omega}_{\phi} & = & \left\{2+\left[3\left(\gamma +
\frac{\pi}{\rho}\right) -2\right] \Omega -
3 \gamma_{\phi}\right\} H \Omega_{\phi}.
\label{eq:generalize}
\end{eqnarray}
\\
We now may have $\dot{\Omega}_{\phi} < 0$ as well as $\Omega_{m}
\rightarrow \Omega_{m0} \neq 0$ and $\Omega_{\phi} \rightarrow
\Omega_{\phi 0} \neq 0$ for late time so long as the stationary
condition
\\
\begin{equation}
\gamma_{m} + \frac{\pi}{\rho_{m}} = \gamma_{\phi} = - \frac{2\dot{H}}
{3H^{2}}
\label{eq:attractor}
\end{equation}
\\
is satisfied. Besides, the constraint $\gamma < 2/3$ is replaced by
$\gamma + (\pi/\rho) < 2/3$, which is somewhat easier to fulfill since
the second law of thermodynamics implies that $\pi$ must  be
negative for expanding fluids (see e.g., Ref. \cite{landau}).

For spatially flat FRW universes, the asymptotic stability of the
stationary solution $\Omega_{m0}$ and $\Omega_{\phi 0}$ can be
studied from equation (\ref{eq:generalize}). By slightly perturbing
$\Omega_{\phi}$ it follows that the solution is stable (and therefore
an attractor) provided the quantity
$\gamma_{m} + \frac{\pi}{\rho_{m}}- \gamma_{\phi}< 0 $ and tends to
zero as $t \rightarrow \infty$. This coincides with the stationary
condition (\ref{eq:attractor}).

For $\Omega \neq 1$ (i.e., when $k \neq 0 $), it is expedient to introduce
the ansatz $\epsilon = \epsilon_{0} + \delta$ in Eqs. (4) \& (5), where
$\epsilon_{0} \equiv (\Omega_{m}/\Omega_{\phi})_{0} \sim {\cal O}(1)$
and $\mid \delta \mid \ll \epsilon_{0}$. One finds that
\\
\begin{equation}
\dot{\delta} = - \frac{3}{\Omega_{\phi}}\left(\frac{2}{3} - \gamma_{\phi}
\right) \Omega_{k} H (\epsilon_{0} + \delta).
\label{delta}
\end{equation}
\\
As a consequence, the stationary solution will be stable for open FRW
universes ($\Omega_{k} > 0$). For closed FRW universes one has to go
beyond the linear analysis.

A realization of these ideas is offered in Ref. \cite{enlarged}. There it is
seen that the space parameter is ample enough that no fine tuning is
required to have late acceleration together with the fact that both
density parameters tend to constant values compatible with
observation.
Moreover, it is well known that for a wide class of dissipative dark
energy models the attractor solutions are themselves attracted
towards a common asymptotic behavior. This ``superattractor'' regime
provides a model of the recent universe that also exhibits an
excellent fit to the high redshift supernovae data luminosity and no
age conflict \cite{qsa}.

Soon after our proof was published some other mechanisms to provide late
cosmic acceleration were proposed. Here we mention two: (i) the
Chaplygin gas \cite{chaplin} and (ii) tachyonic matter \cite{sen}.
\  \\
\  \\
(i) The Chaplygin gas corresponds to a fluid with equation
of state given by
\\
\begin{equation}
p_{ch} = - \frac{A}{\rho_{ch}},
\label{chaplyn}
\end{equation}
\\
where $A$ is a positive--definite constant. This equation has the attractive
features of providing a negative pressure and a speed of sound
always real and positive -something not shared by
quintessence fields.  Support for this exotic component comes from
higher dimensional theories \cite{jackiw}. Likewise, Bento {\it et al.}
demonstrated that Eq. (\ref{chaplyn}) can be derived from a Lagrangian
density of Born--Infeld type \cite{bento}.

Assuming that this fluid does not interact with any other component it
follows that its energy density evolves as
\\
\begin{equation}
\rho_{ch} = \sqrt{A + \frac{B}{a^{6}}},
\label{dependence}
\end{equation}
\\
with $B$ a constant. This dependence has the appealing feature of
leading to $\rho_{ch} \propto a^{-3}$ at early times (dust--type
behavior), and $\rho_{ch} = - p_{ch} = \sqrt{A}$ at late times
(cosmological constant--like behavior). It is obvious, however, that
a universe filled with just this gas, or combined with a perfect fluid
dark matter component, should rely upon fine tuning to solve the
coincidence problem and start accelerating at low redshift.

With the help of Eq. (\ref{dependence}) it is immediately seen that
the equation of state (\ref{chaplyn}) can be written as
$p_{ch} = (\gamma_{ch} - 1) \rho_{ch}$ where
\\
\begin{equation}
\gamma_{ch} = \frac{B}{B + A\, a^{6}}
\label{gammac}
\end{equation}
\\
lies in the range $[0,1]$. As a consequence, the same argument used in
Ref. \cite{enlarged} regarding the coincidence problem when the dark component
was a quintessential scalar field also applies when the latter is
replaced by the Chaplygin gas.\\

At late time, the dynamics is governed by attractor condition
(\ref{eq:attractor}). So, from Eqs. (\ref{eq:attractor}) and (\ref{gammac})
it follows that
\\
\be
\n{at}
\frac{B}{B+Aa^6}=-\frac{2\dot H}{3H^2}
\ee
\\
\no
which for $1 \ll A\, a^{6}/B$ yields the expansion rate
\\
\be
\n{H}
H^2 \simeq H_{0}^{2} {\mbox e}^{B/2Aa^6} \,,
\ee
\\
where $H_{0}$ is an integration constant. In this regime, for $B>0$,
the Chaplygin gas leads to a de Sitter phase and
the energy density of the gas behaves as $\rho\approx\sqrt{A}$. Then
from the Friedmann equation one has $3H_0^2\approx\rho_m+\sqrt{A}$,
while the attractor condition (\ref{eq:attractor}) gives the viscous
pressure $\pi\approx -\gamma_m\rho_m=-\gamma_m(3H_0^2-\sqrt{A})$. For
$B<0$  Eq. (\ref{eq:attractor}) implies that $\dot H>0$,
so the Chaplygin gas gives rise to a superaccelerated expansion.\\

It has been argued that this exotic fluid not only plays the role of dark
energy but also makes non--baryonic dark matter redundant in the sense that
dark matter and dark energy would just be different manifestations of the
Chaplygin gas \cite{bilic}. Thus, one may think that under such circumstances
this scenario evades the coincidence problem. However, this is not the case.
In this unified scenario the coincidence problem is only slightly alleviated.
Indeed assuming a spatially flat universe, the  non--baryonic dark component
would account for about ninety six percent of the critical density and the
baryonic matter (luminous and non--luminous) would account for about four
percent. While these figures are not of exactly the same order, they are not
so different either. They may be seen as nearly coincident, especially if one
bears in mind that at the present time one expects -in view of Eq.
(\ref{dependence})- the dark energy component to be nearly constant while the
baryonic matter redshifts as $a^{-3}$.
\  \\
\  \\

(ii) The tachyonic matter was introduced by Sen \cite{sen} and soon after
its cosmological consequences were explored -see e.g., Ref. \cite{explored}.
In particular Feinstein showed that a never--ending power law inflation
may be achieved provided the tachyonic potential were given by an inverse
square law, $V(\varphi) \propto 1/\varphi^{2}$ \cite{alex}. Obviously
this toy model may also serve for the purpose of late acceleration.

The effective tachyonic fluid is described by the Lagrangian
${\cal L} = - V(\varphi) \sqrt{1 - \partial_{a}\varphi \,
\partial^{a}\varphi }$ where the equation of motion
of the field $\varphi$ in a FRW background takes the form
\\
\[
\ddot{\varphi} +
3H\,\dot{\varphi}\left(1 - \dot{\varphi}^{2}\right)+\frac{\mbox{d}V(\varphi)}{
\mbox{d}\varphi}\frac{{1 - \dot{\varphi}^{2}}}{V(\varphi)}=0.
\]
\\
The corresponding energy density and pressure are given by
\\
\begin{equation}
\rho_{\varphi} = \frac{V(\varphi)}{\sqrt{1 - \dot{\varphi}^{2}}}
\quad \mbox{and} \quad
\quad p_{\varphi} = - V(\varphi) \sqrt{1 - \dot{\varphi}^{2}},
\label{corresponding}
\end{equation}
\\
respectively. They are linked by
$p_{\varphi} = (\gamma_{\varphi} - 1) \rho_{\varphi}$ where
$\gamma_{\varphi} = \dot{\varphi}^{2}$ and is limited to the
interval [0,1]. Again, as in the Chaplygin gas case, the proof
sketched above regarding the quintessence field also applies
when the dark component is tachyonic matter, irrespective
of its potential.

If at late time the scale factor obeys a power--law evolution $a(t) \propto
t^{\alpha}$, then  the attractor condition (\ref{eq:attractor}) -with the
subscript $\phi$ replaced by $\varphi$ - implies that
$\alpha=2/3\gamma_{\varphi}$. Since for tachyonic matter the adiabatic index
is just $\dot{\varphi}^{2}$, we get \cite{padmanabhan}
\\
\be
\varphi(t)=\left(\frac{2}{3\alpha}\right)^{1/2}\,t+\varphi_{0},
\label{c}
\ee
\\
as well as
\\
\be
\n{pot}
V(\varphi)=2\alpha\left(1-\frac{2}{3\alpha}
\right)^{1/2}(\varphi-\varphi_{0})^{-2} \, ,
\ee
\\
\no
where $\alpha>2/3$.
Also, from Eq. (\ref{eq:attractor}), one can obtain the viscous pressure
in the late regime
\\
\be
\pi= \left(\frac{2}{3\alpha}-\gamma_m\right)\rho_{m} <0.
\label{va}
\ee

In summary, the proof offered in Ref. \cite{enlarged} naturally extends
itself to two other dark energy candidates, namely the Chaplygin gas
and the tachyonic effective fluid. Again, in both cases the
dissipative pressure follows by invoking the attractor condition. If
future observations come to show any of them (quintessence, Chaplygin
gas or tachyonic matter) as the right answer to the accelerated
expansion, it could be viewed as a strong indirect support for the
existence of a large dissipative pressure at cosmic scales.

\section*{Acknowledgments} This work has been partially supported by
the Spanish Ministry of Science and Technology under grant BFM
2000-C-03-01 and 2000-1322, and the University of Buenos
Aires under Project X223.


\begin{thebibliography}{99}
\bibitem{enlarged}
L.P. Chimento, A.S. Jakubi and D. Pav\'on, Phys.
Rev. D. {\bf 62}, 063508 (2000).

\bibitem{saul}
S. Perlmutter, Int. J. Mod. Phys. A {\bf 15} Suppl. 1B,
715 (2000).

\bibitem{peacock}
J.A. Peacock, {\it Cosmological Physics} (Cambridge
University Press, Cambridge, 1999).

\bibitem{iap}
Proceedings of the I.A.P. Conference ``On the Nature of Dark
Energy" held in Paris (1-5 July 2002) edit. P. Brax, J. Martin,
and J.P. Uzan (in the press).

\bibitem{quintessence}
C. Wetterich, Nucl. Phys. {\bf B302}, 668 (1988);
B. Ratra and P.E.J. Peebles, Phys. Rev. D {\bf 37}, 3406 (1988);
R.R. Caldwell, R. Dave, and P.J. Steinhardt,  Phys. Rev. Lett. {\bf 80}
1582 (1998);
I. Zlatev, L. Wang, and P.J. Steinhardt, Phys. Rev. Lett. {\bf 82}, 896 (1999).

\bibitem{padmanabhan}
T. Padmanabhan, Phys. Rev. D {\bf 66}, 021301(R) (2002).

\bibitem{landau}
L. Landau and E.M. Lifshitz, {\it M\'ecanique des Fluids} (Editions
MIR, Moscou, 1967).

\bibitem{qsa}
L. P. Chimento, A.S. Jakubi, and N. A. Zuccal\'a, Phys. Rev. D
{\bf 63}, 103508 (2001).

\bibitem{chaplin}
A. Kamenshchik, U. Moschella, and V. Pasquier, Phys.Lett. B. {\bf 511},
265 (2001).

\bibitem{sen}
A. Sen, Mod. Phys. Lett. A {\bf 17}, 1797 (2002);
preprint {\sf [hep-th/0203211]}; preprint {\sf [hep-th/0203265]}.

\bibitem{jackiw}
R. Jackiw, ``(A particle Field Theorist's) Lecture on (Supersymmetric,
Non-Abelian) Fluid Mechanics (and {\it d}-Branes"), preprint
{\sf [physics/0010042]}.

\bibitem{bento}
M.C. Bento, O. Bertolami, and A.A. Sen, Phys. Rev. D {\bf 66}, 043507.

\bibitem{bilic}
N. Bilic, G.B. Tupper, and R.D. Viollier, Phys. Lett. B {\bf 535}, 17 (2002).


\bibitem{explored}
G.W. Gibbons, Phys. Lett B {\bf 537}, 1 (2002);
S. Mukohyama, Phys. Rev D {\bf 66}, 024009 (2002).

\bibitem{alex}
A. Feinstein, Phys. Rev. D {\bf 66}, 063511 (2002).

\end{thebibliography}
\end{document}